\documentclass{article}
\usepackage{latexsym}
\usepackage{amssymb}
\usepackage{amsmath}
\usepackage{amsfonts}
\usepackage{amsthm}
\usepackage{txfonts}
\usepackage{url}
\usepackage[dvips]{graphicx}

\newtheorem{theorem}{Theorem}[section]
\newtheorem{lemma}[theorem]{Lemma}

\newtheorem{rem}[theorem]{Remark}

\newtheorem{definition}[theorem]{Definition}

\newcommand{\2}{\vspace{0.2 cm}}

\newcommand{\ML}[3]{\ell_{\max}(#1,#2,#3)}
\newcommand{\LT}[1]{\mbox{Leaf}(#1)}
\newcommand{\IN}[1]{\mbox{Int}(#1)}
\newcommand{\GrowTree}[1]{T_{D,L}^{root}(#1)}
\newcommand{\GrowTreeT}[2]{#1_{D,L}^{root}(#2)}
\newcommand{\induce}[2]{#1 [ #2 ]}

\title{FPT Algorithms and Kernels for the Directed $k$-Leaf Problem}

\author{Jean Daligault\thanks{Universit\'e Montpellier II, LIRMM, 161 rue
Ada, 34392 Montpellier Cedex 5 - France, {\tt daligault@lirmm.fr}}, Gregory Gutin\thanks{Department of Computer Science, Royal Holloway, University of London,
Egham, Surrey TW20 0EX, UK, {\tt gutin@cs.rhul.ac.uk}}, Eun Jung Kim\thanks{Department of Computer Science,
 Royal Holloway, University of London,
Egham, Surrey TW20 0EX, UK, {\tt eunjung@cs.rhul.ac.uk}}, and Anders Yeo\thanks{Department of Computer Science,
 Royal Holloway, University of London,
Egham, Surrey TW20 0EX, UK, {\tt anders@cs.rhul.ac.uk}}}
\begin{document}
\maketitle

\begin{abstract}
A subgraph $T$ of a digraph $D$ is an {\em out-branching} if $T$ is an
oriented spanning tree with only one vertex
of in-degree zero (called the {\em root}). The vertices of $T$ of
out-degree zero
are {\em leaves}. In the {\sc Directed Max Leaf} Problem, we wish to find the maximum number of leaves in an out-branching of a given
digraph $D$ (or, to report that $D$ has no out-branching).
In the {\sc Directed $k$-Leaf} Problem, we are given a
digraph $D$ and
an integral parameter $k$, and we are to decide whether $D$ has an
out-branching with at least $k$ leaves.
Recently, Kneis et al. (2008) obtained an algorithm for {\sc Directed $k$-Leaf} of
running time $4^{k}\cdot n^{O(1)}$.
We describe a new algorithm for {\sc Directed $k$-Leaf} of running time
$3.72^{k}\cdot n^{O(1)}$. This algorithms leads to an $O(1.9973^n)$-time algorithm
for solving {\sc Directed Max Leaf} on a digraph of order $n.$
The latter algorithm is the first algorithm of running time $O(\gamma^n)$ for {\sc Directed Max Leaf}, where $\gamma<2.$
In the {\sc Rooted Directed $k$-Leaf} Problem,
apart from $D$ and $k$, we are given a vertex $r$ of $D$ and we are to
decide whether $D$ has an out-branching rooted at $r$ with at least $k$
leaves.
Very recently, Fernau et al. (2008) found an $O(k^3)$-size kernel for {\sc
Rooted Directed $k$-Leaf}. In this paper,
we obtain an $O(k)$ kernel for {\sc Rooted Directed $k$-Leaf} restricted
to acyclic digraphs.
\end{abstract}

\section{Introduction}

The {\sc Maximum Leaf} problem is to find a
spanning tree with the maximum number of leaves in a given undirected graph $G.$ The problem is well
studied  from both algorithmic \cite{fominA52,galbiatiTCS181,luJA29,solisLNCS1461} and graph-theoretical \cite{dingJGT3,griggsDM104,kleitmanSIAMJDM4,linial1987} points of view.
The problem has been studied from the parameterized complexity perspective as well and several authors \cite{bonsmaLNCS2747,estivillACID05,fellowsLNCS1974} have designed fixed parameter tractable (FPT) algorithms for solving the parameterized version of {\sc Maximum Leaf} (the {\sc $k$-Leaf} problem): given a graph $G$ and an integral parameter $k$, decide whether $G$ has a spanning tree with at least $k$ leaves.

{\sc Maximum Leaf} can be extended  to digraphs. A subgraph $T$
of a digraph $D$ is an {\em out-tree} if $T$ is an oriented tree with only one vertex
of in-degree zero (called the {\em root}). The vertices of $T$ of out-degree zero
are {\em leaves}. If  $V (T) = V (D)$, then $T$ is an {\em out-branching} of $D$. The {\sc Directed
Maximum Leaf} problem is to find an out-branching with
the maximum number of leaves in an input digraph.
The parameterized version of the {\sc Directed Maximum Leaf} problem
is {\sc Directed $k$-Leaf}:  given a digraph $D$ and
an integral parameter $k$, decide whether $D$ has an out-branching with at least $k$ leaves.
If we add a condition that every out-branching in {\sc Directed $k$-Leaf} must be rooted at a given vertex $r$,
we obtain a variation of {\sc Directed $k$-Leaf} called the {\sc Rooted Directed $k$-Leaf} problem.

The study of {\sc Directed $k$-Leaf}  has only begun recently.
Alon et al. \cite{alonLNCS4596,alonLNCS4855} proved that the problem is FPT for a wide family of digraphs including classes of strongly connected and acyclic digraphs. Bonsma and Dorn  extended this result to all digraphs in \cite{bonsma2007}, and improved
the running time of the algorithm in \cite{alonLNCS4855} to  $2^{k\log k}n^{O(1)}$ in \cite{bonsmaESA}. Recently, Kneis et al. \cite{kneisISAAC2008} obtained an
algorithm for solving the problem in time $4^kn^{O(1)}$. Notice that the algorithm of Kneis et al. \cite{kneisISAAC2008} applied to undirected graphs is of smaller running time (as a function of $k$) than all previously known algorithms for {\sc $k$-Leaf}. Yet, the algorithm of Kneis et al. \cite{kneisISAAC2008} is not fast enough to answer in affirmative the question of Fellows et al. \cite{fellowsLNCS1974} of whether there exists a parameterized algorithm for {\sc Max Leaf} of running time
$f(k)n^{O(1)}$, where $f(50)<10^{20}.$
Very recently, Fernau et al. \cite{fernau} proved that no polynomial kernel for {\sc Directed $k$-Leaf} is possible unless the polynomial hierarchy collapses to the third level (they applied a recent breakthrough result of Bodlaender et al. \cite{bodlaenderLNCS5125}). Interestingly, {\sc Rooted Directed $k$-Leaf} admits a polynomial size kernel and  Fernau et al. \cite{fernau} obtained one of size $O(k^3).$

The only known approximation algorithm for {\sc Directed Max Leaf} is due to Drescher and Vetta \cite{drescherACMTA} and its approximation ratio is
$O(\sqrt{\ell_{\max}(D)})$, where $\ell_{\max}(D)$ is the maximum number of leaves in an out-branching of a digraph $D$.

In this paper, we obtain an algorithm faster than the one of Kneis et al. \cite{kneisISAAC2008} for {\sc Directed $k$-Leaf}. Our algorithm runs in time $3.72^kn^{O(1)}.$ Unfortunately, our algorithm cannot solve the above-mentioned question of Fellows et al. \cite{fellowsLNCS1974}, but it shows that the remaining gap is not wide anymore.
We also obtain a linear size kernel for {\sc Directed $k$-Leaf}
restricted to acyclic digraphs. Notice that (i) {\sc Directed Max Leaf} restricted to acyclic digraphs is still NP-hard \cite{alonSIAMJDM}, and (ii)
for acyclic digraphs {\sc Directed $k$-Leaf} and {\sc Rooted Directed $k$-Leaf} are equivalent since all out-branchings must be rooted at the unique vertex of in-degree zero.

We recall some basic notions of parameterized complexity here, for a
more in-depth treatment of the topic we refer the reader to
\cite{downey1999,flum2006,niedermeier2006}.

A parameterized problem $\Pi$ can be considered as a set of pairs
$(I,k)$ where $I$ is the \emph{problem instance} and $k$ (usually an
integer) is the \emph{parameter}.  $\Pi$ is called
\emph{fixed-parameter tractable (FPT)} if membership of $(I,k)$ in
$\Pi$ can be decided in time $O(f(k)|I|^c)$, where $|I|$ is the size
of $I$, $f(k)$ is a computable function, and $c$ is a constant
independent from $k$ and $I$. Let $\Pi$ be a parameterized problem. A \emph{reduction $R$ to a problem kernel} (or \emph{kernelization}) is a many-to-one transformation from $(I,k)\in\Pi$ to $(I',k')\in \Pi'$, such that (i) $(I,k)\in \Pi$ if and only if $(I',k')\in \Pi$, (ii) $k'\le k$ and $|I'|\le g(k)$ for some function $g$ and (iii) $R$ is computable in time polynomial in $|I|$ and $k$. In kernelization, an instance $(I,k)$ is
reduced to another instance $(I',k')$, which is called the
\emph{problem kernel}; $|I'|$ is the {\em size} of the kernel.

The set of vertices (arcs) of a digraph $D$ will be denoted by $V(D)$ ($A(D)$). The number of vertices (arcs) of the digraph under consideration will be denoted $n$ ($m$).
For a vertex $x$ of a subgraph $H$ of a digraph $D$, $N_H^+(x)$ and $N_H^-(x)$ denote the sets of
out-neighbors and in-neighbors of $x$, respectively. Also, let $A_H^+(x)=\{xy:\ y\in N^+_H(x)\},$ $d^+_H(x)=|N^+_H(x)|$, and $d^-_H(x)=|N^-_H(x)|.$
When $H=D$ we will omit the subscripts in the notation above.

Let $D$ be a digraph, $T$ an out-tree and $L\subseteq V(D)$. A $(T,L)$-\emph{out-tree} of $D$ is an out-tree $T'$ of $D$ such that (1) $A(T) \subseteq A(T')$, (2) $L$ are leaves in $T'$, (3) $T$ and $T'$ have the same root. A $(T,L)$-\emph{out-branching} is a $(T,L)$-out-tree which is spanning. Let $\ML{D}{T}{L}$ be the maximum number of leaves over all $(T,L)$-out-branchings of $D$. We set this number to 0 if there is no $(T,L)$-out-branching. For an out-tree $T$ in a digraph $D$, $\LT{T}$ denotes the set of leaves in $T$ and $\IN{T}=V(T)-\LT{T}$, the set of {\em internal vertices} of $T$. For any vertex $x$ in a tree $T$ let $T_x$ denote the maximal subtree of $T$ which has $x$ as its root.

Throughout this paper we use a triple $(D,T,L)$ to denote a given digraph $D$, an out-tree $T$ of $D$ and a set of vertices $L \subseteq V(D)-\IN{T}$. We denote by $\hat{D}(T,L)$ the subgraph of $D$ obtained after deleting all arcs out of vertices in $L$ and all arcs not in $A(T)$ which go into a vertex in $V(T)$. When $T$ and $L$ are clear from the context we will omit them and denote $\hat{D}(T,L)$ by $\hat{D}$. For further terminology and notation
on directed graphs, one may consult \cite{bang2000}. The following simple lemma will be used in the rest of the paper.

\begin{lemma}\label{OBexist}\cite{bang2000}
A digraph $D$ has an
out-branching if and only if $D$ has a single strong component
without incoming arcs. One can decide whether a digraph has an out-branching in time $O(n+m).$
\end{lemma}

\section{Another $4^kn^{O(1)}$ Time Algorithm}

The algorithm of this section is similar to the algorithm in \cite{kneisISAAC2008}, but it differs from the algorithm in \cite{kneisISAAC2008} as follows. We decide in an earlier stage which one of the current leaves of $T$ cannot be a leaf in a final $(T,L)$-out-branching and make them to be internal vertices based on Lemma \ref{branch}, see step 2 in Algorithm $\mathcal{A}(D,T,L)$. This decision works as a preprocessing of the given instance and gives us a better chance to come up with a $(T,L)$-out-tree with at least $k$ leaves more quickly. A more important reason for this step is the fact that our algorithm is easier than the main algorithm in \cite{kneisISAAC2008} to transform into a faster algorithm.

The following simple result was used in \cite{alonLNCS4596, alonLNCS4855} and its proof can be found in \cite{kneisISAAC2008}.

\begin{lemma}\label{extension}
If there is an out-branching rooted at vertex $r$, whenever we have an out-tree rooted at $r$ with at least $k$ leaves we can extend it to an out-branching rooted at $r$ with at least $k$ leaves in time $O(m+n)$.
\end{lemma}

\begin{lemma}\label{redgreen}
Given a triple $(D,T,L)$,
 we have $\ML{D}{T}{L} = \ML{\hat{D}}{T}{L}$.
\end{lemma}
\begin{proof}
If there is no $(T,L)$-out-branching in $D$, the subgraph $\hat{D}$ does not have a $(T,L)$-out-branching either and the equality holds trivially. Hence suppose that $T^*$ is a $(T,L)$-out-branching in $D$ with $\ML{D}{T}{L}$ leaves. Obviously we have $\ML{D}{T}{L} \geq \ML{\hat{D}}{T}{L}$. Since the vertices of $L$ are leaves in $T^*$, all arcs out of vertices in $L$ do not appear in $T^*$, i.e. $A(T^*)\subseteq A(D) \setminus \{A^+_D(x):x\in L\}$. Moreover $A(T)\subseteq A(T^*)$ and thus all arcs not in $A(T)$ which go into a vertex in $V(T)$ do not appear in $T^*$ since otherwise we have a vertex in $V(T)$ with more than one arc of $T^*$ going into it (or, the root has an arc going into it). Hence we have $A(T^*)\subseteq A(\hat{D})$ and the above equality holds.
\end{proof}

\begin{lemma}\label{branch}
Given a triple $(D,T,L)$, the following equality holds for each leaf $x$ of $T$.
$$\ML{D}{T}{L}  = \max\{ \ML{D}{T}{L\cup \{x\}}, \ML{D}{T\cup A^+_{\hat{D}}(x)}{L}\}$$

\end{lemma}
\begin{proof}
If $\ML{D}{T}{L}=0$ then the equality trivially holds, so we assume that $\ML{D}{T}{L}\geq 1$. Since any $(T,L\cup \{x\})$-out-branching or $(T\cup A^+_{\hat{D}}(x),L)$-out-branching is a
$(T,L)$-out-branching as well, the inequality $\geq$ obviously holds.
To show the opposite direction, suppose $T'$ is an optimal $(T,L)$-out-branching.
If $x$ is a leaf in $T'$, then $T'$ is a $(T,L\cup \{x\})$-out-branching and $\ML{D}{T}{L} \leq \ML{D}{T}{L\cup\{x\}}$.

Suppose $x$ is not a leaf in $T'$. Delete all arcs entering $N^+_{\hat{D}}(x)$ in $T'$, add $A^+_{\hat{D}}(x)$
and let $T''$ denote the resulting subgraph. Note that $d^-_{T''}(y)=1$ for each vertex
$y$ in $T''$ which is not the root and $A(T'')\subseteq A(\hat{D})$. In order to show that $T''$ is an out-branching
it suffices to see that there is no cycle in $T''$ containing $x$.
If there is a cycle $C$ containing $x$ in $T''$ and $xy\in A(C)$,
then $C-\{xy\}$ forms a directed $(y,x)$-path in $\hat{D}$. However this is a contradiction as
$x \in V(T)$ and $y \not\in V(T)$ and there is no path from $V(D)-V(T)$ to $V(T)$ in $\hat{D}$.
Hence $T''$ is an out-branching.

As no vertex in $L$ has any arcs out of it in $\hat{D}$ we note that $L \subseteq \LT{T''}$. Furthermore we note that
$A(T) \subseteq A(T'')$ as $A(T) \subseteq A(T')$ and all arcs we deleted from $A(T')$ go to a vertex not in $V(T)$.
Therefore $T''$ is a $(T,L)$-out-branching which has as many leaves as $T'$.
This shows $\ML{D}{T}{L} \leq \ML{D}{T \cup A^+_{D'}(x)}{L}$.
\end{proof}

\begin{definition} \label{DefGrowTree}
Given a triple $(D,T,L)$ and a vertex $x \in \LT{T} - L$, define $\GrowTree{x}$ as follows.

\begin{quote}
\begin{description}
 \item[(1)] $x' \coloneqq x$.
 \item[(2)] While $d^+_{\hat{D}}(x')=1$
 add $A^+_{\hat{D}}(x')=\{x'y\}$ to $T$ and let $x' \coloneqq y$.
 \item[(3)] Add $A^+_{\hat{D}}(x')$ to $T$.
\end{description}
\end{quote}

Now let $\GrowTree{x}=T_x$. That is, $\GrowTree{x}$ contains exactly the arcs added by the above process.
\end{definition}

The idea behind this definition is the following: during the algorithm, we will decide that a given leaf $x$ of the partial out-tree $T$ built thus far is not a leaf of the out-branching we are looking for. Then adding the out-arcs of $x$ to $T$ is correct. To make sure that the number of leaves of $T$ has increased even when $d^+_{V-V(T)}(x)=1$, we add $T^{root}(x)$ to $T$ instead of just adding the single out-arc of $x$, as described in the following.

\begin{lemma} \label{NewBranch}
Suppose we are given a triple $(D,T,L)$ and a leaf $x \in \LT{T}-L$. If $\ML{D}{T}{L\cup \{x\}} \geq 1$
then the following holds.

\begin{description}
\item[(i)] If $|\LT{\GrowTree{x}}| \geq 2$ then $\ML{D}{T}{L} = \max \{ \ML{D}{T}{L \cup \{x\}}, \ML{D}{T \cup \GrowTree{x}}{L}$.
\item[(ii)] If $|\LT{\GrowTree{x}}| = 1$  then $\ML{D}{T}{L} = \ML{D}{T}{L \cup \{x\}}$.
\end{description}
\end{lemma}

\begin{proof}
Assume that $T'$ is an optimal $(T,L)$-out-branching and that $|\LT{T_x'}|=1$. We will now show that
$ \ML{D}{T}{L \cup \{x\}} = |\LT{T'}| = \ML{D}{T}{L}$.
If $x$ is a leaf of $T'$ then this is clearly the case, so assume that $x$ is not a leaf of $T'$.
Let $y$ be the unique out-neighbor of $x$ in $T'$. As $\ML{D}{T}{L\cup \{x\}} \geq 1$ we note that there
exists a path $P=p_0p_1p_2 \ldots p_r(= y)$ from the root of $T$ to $y$ in $\hat{D}(T,L \cup \{x\})$.  Assume that
$q$ is chosen such that $p_q \not\in T_x'$ and $\{p_{q+1},p_{q+2},\ldots ,p_r\} \subseteq V(T_x')$.
Consider the digraph $D^*=\induce{D}{V(T_x') \cup \{p_q\} -\{x\}}$ and note that $p_q$ can reach all vertices
in $D^*$. Therefore there exists an out-branching in $D^*$, say $T^*$, with $p_q$ as the root. Let $T''$ be
the out-branching obtained from $T'$ by deleting all arcs in $T_x'$ and adding all arcs in $T^*$. Note that
$|\LT{T''}| \geq |\LT{T'}|$ as $\LT{T^*} \cup \{x\}$ are leaves in $T''$ and $\LT{T'_x} \cup \{p_q\}$ are the
only leaves in $T'$ which may not be leaves in $T''$ (and $|\LT{T'_x} \cup \{p_q\}|=2$).  Therefore
$ \ML{D}{T}{L \cup \{x\}} \geq |\LT{T'}| = \ML{D}{T}{L}$. As we always have $\ML{D}{T}{L} \geq \ML{D}{T}{L \cup \{x\}}$ we get the desired equality.

This proves part (ii) of the lemma, as if $|\LT{\GrowTree{x}}| = 1$ then any  optimal $(T,L)$-out-branching $T'$,
must have $|\LT{T_x'}|=1$.

We therefore consider part (i), where $|\LT{\GrowTree{x}}| \geq 2$. Let $Q$ denote the set of leaves of $\GrowTree{x}$ and let $R=V(\GrowTree{x})-Q$. Note that by the construction of $\GrowTree{x}$ the vertices of $R$ can be ordered $(x=)r_1,r_2,\ldots,r_i$ such that $r_1r_2\ldots ,r_i$ is a path in
$\GrowTree{x}$. As before let $T'$ be an optimal $(T,L)$-out-branching and note that if any
$r_j$ ($1 \leq j \leq i$) is a leaf of $T'$ then $|\LT{T_x'}|=1$ and the above gives us
$ \ML{D}{T}{L \cup \{x\}}  = \ML{D}{T}{L}$. This proves part (i) in this case, as we always have
$\ML{D}{T}{L} \geq \ML{D}{T \cup \GrowTree{x}}{L}$. Therefore no vertex in $\{r_1,r_2,\ldots,r_i\}$
is a leaf of $T'$ and all arcs $(x=)r_1r_2, r_2r_3, \ldots ,r_{i-1}r_i$ belong to $T'$.
By Lemma \ref{branch} we may furthermore assume that $T'$ contains all the arcs from $r_i$ to vertices
in $Q$. Therefore $\GrowTree{x}$ is a subtree of $T'$ and $\ML{D}{T}{L} = \ML{D}{T \cup \GrowTree{x}}{L}$.
This completes the proof of part (i).
\end{proof}

\noindent The following is an $O(4^kn^{O(1)})$ algorithm. Its complexity can be obtained similarly to \cite{kneisISAAC2008}. We restrict ourselves only to proving its correctness.

\begin{quote}
For every vertex $x \in V(D)$, do $\mathcal{A}(D,\{x\},\emptyset)$.

If one of the returns of $\mathcal{A}(D,\{x\},\emptyset)$ is ``YES'' then output ``YES''.

Otherwise, output ``NO''.

\vspace{20pt}

\noindent{$\mathcal{A}(D,T,L)$:}
\begin{description}
 \item[(1)]  If $\ML{D}{T}{L}=0$, return ``NO''. Stop.
 \item[(2)]  While there is a vertex $x \in \LT{T}-L$ such that $\ML{D}{T}{L\cup \{x\}} = 0$, add the arcs $A^+_{\hat{D}}(x)$ to $T$.
 \item[(3)]
          If $|L| \geq k$, return ``YES''. Stop.
 \newline If the number of leaves in $T$ is at least $k$, return ``YES''. Stop.
 \newline If all leaves in $T$ belong to $L$, return ``NO''. Stop.
 \item[(4)] Choose a vertex $x \in \LT{T}-L$.
 \newline $B_1:=\mathcal{A}(D,T,L\cup\{x\})$ and $B_2:=$``NO''.
 \newline If $|\LT{\GrowTree{x}}| \geq 2$ then let $B_2:=\mathcal{A}(D,T \cup \GrowTree{x},L)$.
 \newline Return ``YES'' if either $B_1$ or $B_2$ is ``YES''. Otherwise return ``NO''.
\end{description}
\end{quote}

\begin{rem}
While the first line in step 3 is unnecessary, we keep it since it is needed in the next algorithm where $L \subseteq \LT{T}$ is not necessarily true, see (4.2) in the next algorithm, where $p_0\not\in V(T)$.
\end{rem}

\begin{theorem}\label{slow}
Algorithm $\mathcal{A}(D,T,L)$ works correctly. In other words, $D$ has a $(T,L)$-out-branching with at
least $k$ leaves if and only if Algorithm $\mathcal{A}(D,T,L)$ returns ``YES''.
\end{theorem}
\begin{proof}
We begin by showing that a call to $\mathcal{A}(D,T,L)$ is always made with a proper argument $(D,T,L)$,
that is, $T$ is an out-tree of $D$ and $L\cap \IN{T}=\emptyset$. Obviously the initial
argument $(D,\{x\}, \emptyset)$ is proper. Suppose $(D,T,L)$ is a proper argument. It is easy
to see that $(D,T,L\cup\{x\})$ is a proper argument. Let us consider $(D,T \cup \GrowTree{x},L)$.
By Definition \ref{DefGrowTree} we note that $T \cup \GrowTree{x}$ is an out-tree in $D$ and since we
consider the digraph $\hat{D}$ at each step in Definition \ref{DefGrowTree} we note that no vertex in $L$ is
an internal vertex of $T \cup \GrowTree{x}$.
Hence $(D,T \cup \GrowTree{x},L)$ is a proper argument.

Consider the search tree $ST$ that we obtain by running the algorithm
$\mathcal{A}(D,T,L)$. First consider the case when $ST$ consists of a single node. If
$\mathcal{A}(D,T,L)$ returns "NO" in step 1, then clearly we do not have a
$(T,L)$-out-branching. Step 2 is valid by Lemma \ref{branch}, i.e. it does
not change the return of $\mathcal{A}(D,T,L)$. So now consider Step 3. As
$\ML{D}{T}{L} \geq 1$ after step 1, and by Lemma \ref{branch} the value of
$\ML{D}{T}{L}$ does not change by step 2 we note that $\ML{D}{T}{L} \geq
1$ before we perform step 3. Therefore there exists a
$(T,L)$-out-branching in $D$. If $|L| \geq k$ or $|\LT{T}| \geq k$ then,
by Lemma \ref{extension}, any $(T,L)$-out-branching in $D$ has at least
$k$ leaves and
the algorithm returns ``YES''.  If $\LT{T} \subseteq L$ then the only
$(T,L)$-out-branching in $D$ is $T$ itself
and as $|\LT{T}| < k$ the algorithm returns ``NO'' as it must do. Thus, the theorem holds when $ST$ is just a node.

Now suppose that $ST$ has at least two nodes and the theorem holds for all
successors of the root $R$ of $ST$. By the assumption that $R$ makes
further recursive calls, we have $\ML{D}{T}{L} \geq 1$ and there exists a
vertex $x\in \LT{T}-L$. If there is a $(T,L)$-out-branching with at least
$k$ leaves, then by Lemma \ref{NewBranch} there is a
$(T,L\cup\{x\})$-out-branching with at least $k$ leaves or $(T \cup
\GrowTree{x},L)$-out-branching with at least $k$ leaves. By induction
hypothesis, one of $B_1$ or $B_2$ is ``YES'' and thus $\mathcal{A}(D,T,L)$
correctly returns "YES". Else if $\ML{D}{T}{L} < k$, then again by Lemma
\ref{NewBranch} and induction hypothesis both $B_1$ and $B_2$ are "NO".
Therefore the theorem holds for the root $R$ of $ST$, which completes the proof.
\end{proof}

\section{Faster Algorithm}

We now show how the algorithm from the previous section can be made faster by
adding an extra vertex to the set $L$ in certain circumstances.
Recall that Step 2 in the above algorithm $\mathcal{A}(D,T,L)$ and in our new algorithm $\mathcal{B}(D,T,L)$ is new
compared to the algorithm in \cite{kneisISAAC2008}. We will also  allow
$L$ to contain vertices which are not leaves of the current out-tree $T$.
The improved algorithm is now described.

\begin{quote}

For every vertex $x \in V(D)$, do $\mathcal{B}(D,\{x\},\emptyset)$.

If one of the returns of $\mathcal{B}(D,\{x\},\emptyset)$ is ``YES'' then output ``YES''.

Otherwise, output ``NO''.

\vspace{20pt}

\noindent $\mathcal{B}(D,T,L):$
\begin{description}
 \item[(1)]  If $\ML{D}{T}{L}=0$, return ``NO''. Stop.
 \item[(2)]  While there is a vertex $x \in \LT{T}-L$ such that $\ML{D}{T}{L\cup \{x\}} = 0$, then
             add the arcs $A^+_{\hat{D}}(x)$ to $T$.
 \item[(3)]
          If $|L| \geq k$, return ``YES''. Stop.
 \newline If the number of leaves in $T$ is at least $k$, return ``YES''. Stop.
 \newline If all leaves in $T$ belong to $L$, return ``NO''. Stop.
 \item[(4)] Choose a vertex $x \in \LT{T}-L$, color $x$ red and let $H_x:=\hat{D}$.
\begin{description}
 \item[(4.1)] Let $z$ be the nearest ancestor of $x$ in $T$ colored red, if it exists.
 \item[(4.2)] Let $L':=L \cup \{x\}$.
  \newline If $z$ exists and $T_{z}$ has exactly two leaves $x$ and $x'$ and $x' \in L$ then:
   \newline Let $P=p_0p_1\ldots p_r$ be a path in $H_z-A^+_{\hat{D}}(z)$ such that $V(P)-V(T_z) = \{p_0\}$ and
           $p_r \in N^+_{\hat{D}}(z)$, and let $L':=L \cup \{p_0,x\}$.
 \item[(4.3)] $B_1:=\mathcal{B}(D,T,L')$ and $B_2:=$``NO''.
 \item[(4.4)] If $|\LT{\GrowTree{x}}| \geq 2$ then let $B_2:=\mathcal{B}(D,T \cup \GrowTree{x},L)$.
 \item[(4.5)] Return ``YES'' if either $B_1$ or $B_2$ is ``YES''. Otherwise return ``NO''.
\end{description}
\end{description}
\end{quote}

The existence of $P$ in step (4.2) follows from the fact that $z$ was
colored red, hence adding $z$ to $L$ would not have destroyed all
out-branchings. Note that $p_0$ does not necessarily belong to $T$.

For the sake of simplifying the proof of Theorem~\ref{fastComplexity} below we furthermore assume that the above algorithm picks the vertex $x$ in Step 4 in a depth-first manner. That is, the vertex $x$ is chosen to be the last vertex added to $T$ such that $x \in \LT{T}-L$.

\begin{theorem}\label{fast}
Algorithm $\mathcal{B}(D,T,L)$ works correctly. In other words, $D$ has a $(T,L)$-out-branching with at least $k$ leaves if and only if Algorithm $\mathcal{B}(D,T,L)$ returns ``YES''.
\end{theorem}
\begin{proof}
The only difference between $\mathcal{B}(D,T,L)$ and $\mathcal{A}(D,T,L)$ is that in step (4.2) we may add an extra vertex $p_0$
to $L$ which was not done in $\mathcal{A}(D,T,L)$. We will now prove that this addition does not change the correctness
of the algorithm.

So assume that there is an optimal $(T,L)$-out-branching $T'$ with $x \in \LT{T'}$ but
$p_0 \not\in \LT{T'}$. We will show that this implies that an optimal
solution is found in the branch of the search tree where we put $z$ into $L$.
This will complete the proof as if an optimal $(T,L)$-out-branching $T'$
does not contain $x$ as a leaf, by Lemma \ref{NewBranch}  it is found in
$\mathcal{B}(D,T \cup \GrowTree{x},L)$ and if it includes both $x$ and $p_0$ as
leaves then it is found in $B(D,T,L')$ (in step (4.3)).

Note that $T_z=T'_z$ as $T_{z}$ had exactly two leaves $x$ and $x'$ and $x' \in L$ and we have
just assumed that $x$ is a leaf of $T'$.
Let $D^*=\induce{D}{V(T_z') \cup \{p_0\} -\{z\}}$ and consider the following two cases.

If $p_0$ can reach all vertices of $D^*$ in $D^*$ then proceed as follows.
Let $T^*$ be an  out-branching in $D^*$ with $p_0$ as the root.  Let $T''$ be
the out-branching obtained from $T'$ by deleting all arcs in $T_z'$ and adding all arcs in $T^*$. Note that
$|\LT{T''}| \geq |\LT{T'}|$ as $\LT{T^*} \cup \{z\}$ are leaves in $T''$ and $\LT{T'_z}$ are the
only two leaves in $T'$ which may not be leaves in $T''$. Therefore an optimal solution
is found when we add $z$ to $L$.

So now consider the case when $p_0$ cannot reach all vertices of $D^*$ in $D^*$.
This means that there is a vertex $u \in N^+_{T}(z)$ which cannot be reached by $p_0$ in $D^*$. All such unreachable vertices lie on the same branch of $T_z$ (the branch not containing $p_r$).
Let $W=w_0w_1w_2 \ldots w_lu$ be a path from the root of $T$ to $u$, which does not
use any arcs out of $z$ (which exists as $z$ was colored red in step (4.1), so adding
$z$ to $L$ at this stage would not destroy all out-branchings).
 Assume that
$a$ is chosen such that $w_a \not\in T_z'$ and $\{w_{a+1},w_{a+2},\ldots ,w_l,u\} \subseteq V(T_z')$ (see Figure 1).

\begin{figure}
   \begin{center}
       \includegraphics [scale = .8, angle =0]{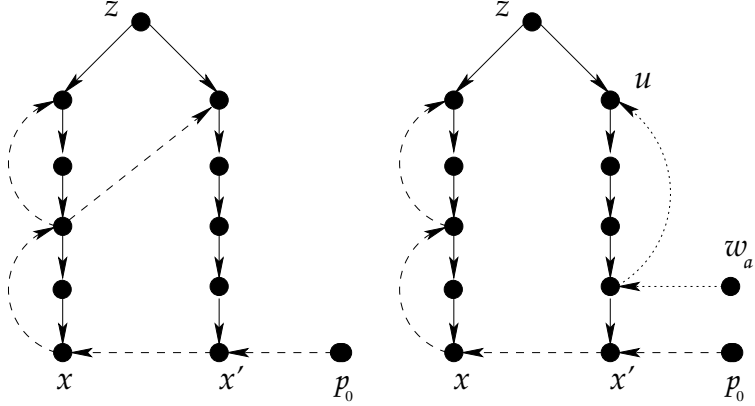}
   \end{center}
   \caption{Real lines represents $T'_z$ arcs; dashed lines represent the reachability of $p_0$; dotted lines represent the reachability of $w_0$.}
   \label{fig1}
\end{figure}
Consider the digraph $D''=\induce{D}{V(T_z') \cup \{p_0,w_a\} -\{z\}}$ and note that every vertex in $D''$ can
be reached by either $p_0$ or $w_a$ in $D''$. Therefore, there exists two vertex disjoint out-trees $T_{p_0}$ and
$T_{w_a}$ rooted at $p_0$ and $w_a$, respectively, such that $V(T_{p_0}) \cup V(T_{w_a}) = V(D'')$
(to see that this claim holds add a new vertex $y$ and two arcs $yp_0$ and $yw_a$).
Furthermore since $p_0$ cannot reach $u$ in $D^*$ we note that both $T_{p_0}$ and
$T_{w_a}$ must contain at least two vertices.
Let $T'''$ be
the out-branching obtained from $T'$ by deleting all arcs in $T_z'$ and adding all arcs in $T_{p_0}$ and in
$T_{w_a}$. Note that
$|\LT{T'''}| \geq |\LT{T'}|$ as $\LT{T_{p_0}} \cup \LT{T_{w_a}} \cup \{z\}$ are leaves in $T'''$ and
$\LT{T'_z} \cup \{w_a\}$ are the
only three vertices which may be leaves in $T'$ but not in $T'''$.  Therefore again an optimal solution
is found when we add $z$ to $L$.
\end{proof}

\begin{theorem}\label{fastComplexity}
Algorithm $\mathcal{B}(D,T,L)$ runs in time $O(3.72^kn^{O(1)})$.
\end{theorem}

\begin{proof}
For an out-tree $Q$, let $\ell(Q)=|Leaf(Q)|$. Recall that we have assumed that $\mathcal{B}(D,T,L)$ picks the vertex $x$ in Step 4 in a depth-first manner.

Consider the search tree $ST$ that we obtain by running the algorithm $\mathcal{B}(D,\{x\},\emptyset)$.
That is, the root of $ST$ is the triple $(D,\{x\},\emptyset)$. The children of this root is
$(D,\{x\},L')$ when we make a recursive call in step (4.3) and $(D,\GrowTree{x},\emptyset)$
if we make a recursive call in step (4.4). The children of these nodes are again triples corresponding to the recursive calls.

Let $g(T,L)$ be the number of leaves in a subtree $R$ of $ST$ with triple $(D,T,L)$. Clearly, $g(T,L)=1$ when $(D,T,L)$ is a leaf of $ST$.
For a non-trivial subtree $R$ of $ST$,
we will prove, by induction, that  $ g(T,L) \le c \alpha^{k-\ell(T)} \beta^{k-|L|} $,
where $\alpha=1.96,\ \beta=1.896$ and $c\geq \alpha^{2} \beta^{2}.$ Assume that this holds for all smaller non-trivial subtrees.
(Note that the value of $c$ is chosen in such a way that in the inequalities in the rest of the proof, we have upper bounds for $g(T^*,L^*)$ being at least 1 when $(D,T^*,L^*)$ is a leaf of $ST$.)

Recall that  $x \in \LT{T}-L$  was picked in step (4). Now consider the following possibilities.

If $|L'| = |L|+2$, then the number of leaves of $R$ is at most the following as if
a call is made to $\mathcal{B}(D,T \cup \GrowTree{x},L)$ in (4.4) then the number of leaves of $T$ increases by at least one:

\2

$\begin{array}{rcl}
g(T,L') + g(T \cup \GrowTree{x},L) & \leq &
c \alpha^{k-\ell(T)} \beta^{k-|L|-2} + c \alpha^{k-\ell(T)-1}  \beta^{k-|L|} \\
  & = & c \alpha^{k-\ell(T)} \beta^{k-|L|} \left( \frac{1}{\beta^2} + \frac{1}{\alpha} \right) \\
  & \leq & c \alpha^{k-\ell(T)} \beta^{k-|L|}. \\
\end{array}$

\2

So we may assume that $|L'| = |L|+1$ in (4.3). Now assume that $|\LT{\GrowTree{x}}| \not= 2$ in (4.4).
In this case either no recursive call is made in (4.4) or we increase the number of leaves in $T$ by at least
two. Therefore the number of leaves of $R$ is at most

\2

$\begin{array}{rcl}
c \alpha^{k-\ell(T)} \beta^{k-|L|-1} + c \alpha^{k-\ell(T)-2}  \beta^{k-|L|}
  & = & c \alpha^{k-\ell(T)} \beta^{k-|L|} \left( \frac{1}{\beta} + \frac{1}{\alpha^2} \right) \\
  & \leq & c \alpha^{k-\ell(T)} \beta^{k-|L|}.\\
\end{array}$

\2

So we may assume that $|L'| = |L|+1$ in (4.3) and $|\LT{\GrowTree{x}}| = 2$ in (4.4).
Let $T' = T \cup \GrowTree{x}$ and consider the recursive call to $\mathcal{B}(D,T',L)$. If we
increase the number of leaves in $T'$ in step (2) of this recursive call, then the number of leaves of the
subtree of $ST$ rooted at $(D,T',L)$ is at most

\2

$\begin{array}{rcl}
c \alpha^{k-\ell(T')-1} \beta^{k-|L|-1} + c \alpha^{k-\ell(T')-2}  \beta^{k-|L|}
  & = & c \alpha^{k-\ell(T')} \beta^{k-|L|} \left( \frac{1}{\alpha \beta} + \frac{1}{\alpha^2} \right). \\
\end{array}$

\2

Therefore, as $\ell(T')=\ell(T)+1$, the number of leaves in $R$  is at most

\2

$\begin{array}{rcl}
g(T,L') + g(T',L) & \leq &
c \alpha^{k-\ell(T)} \beta^{k-|L|-1} + c \alpha^{k-\ell(T)-1}  \beta^{k-|L|} \left(\frac{1}{\alpha \beta} + \frac{1}{\alpha^2} \right) \\
  & = & c \alpha^{k-\ell(T)} \beta^{k-|L|} \left( \frac{1}{\beta} + \frac{1}{\alpha^2 \beta} + \frac{1}{\alpha^3} \right) \\
  & \leq & c \alpha^{k-\ell(T)} \beta^{k-|L|}. \\
\end{array}$

\2

So we may assume that we do not increase the number of leaves in step (2) when we consider
$(D,T',L)$. Let $y$ and $y'$ denote the two leaves of $T'_x$ (after possibly adding some
arcs in step (2)). Consider the recursive call to $\mathcal{B}(D,T',L \cup \{y\})$. If we increase
the number of leaves of $T'$ in step (2) in this call then  the number of leaves in $R$  is at most

\2

$\begin{array}{rcl}
g(T,L \cup \{x\}) & + & g(T', L \cup \{y\}) \; \; + \; \; g(T' \cup \GrowTreeT{(T')}{y},L) \\
& \leq & c \alpha^{k-\ell(T)} \beta^{k-|L|} \left( \frac{1}{\beta} + (\frac{1}{\alpha^2 \beta^2} + \frac{1}{\alpha^3 \beta}) + \frac{1}{\alpha^2} \right) \\
& \leq & c \alpha^{k-\ell(T)} \beta^{k-|L|}. \\
\end{array}$

\2

So we may assume that we do not increase the number of leaves in step (2) when we consider
$(D,T',L \cup \{y\})$.  However in this case we note that $|L'|=|L|+2$ in this recursive call as when we consider $y'$ the
conditions of (4.2) are satisfied as, in particular, $T_x$ has exactly two leaves). So in this last case the number of leaves in $R$  is at most

\2

$\begin{array}{rcl}
g(T,L \cup \{x\})  & + & g(T', L \cup \{y\}) \; \; +\;  \; g(T' \cup \GrowTreeT{(T')}{y},L) \\
& \leq & c \alpha^{k-\ell(T)} \beta^{k-|L|} \left( \frac{1}{\beta} + (\frac{1}{\alpha \beta^3} + \frac{1}{\alpha^2 \beta}) + \frac{1}{\alpha^2} \right)  \\
& \leq & c \alpha^{k-\ell(T)} \beta^{k-|L|}. \\
\end{array}$

\2

We increase either $|L|$ or $\ell(T)$ whenever we consider a child in the search tree and no non-leaf in $ST$ has $|L| \geq k$ or $\ell(T) \geq k$. Therefore, the number of nodes in
$ST$ is at most $O(k \alpha^k  \beta^k) = O(3.72^k)$. As the amount of work we do in each recursive
call is polynomial we get the desired time bound.
\end{proof}

\section{Exponential Algorithm for {\sc Directed Maximum Leaf}}\label{admlsec}
Note that {\sc Directed Maximum Leaf} can be solved in time $O(2^nn^{O(1)})$ by an exhaustive search using Lemma \ref{OBexist}. Our $3.72^kn^{O(1)}$ algorithm for {\sc Directed $k$-Leaf} yields an improvement for {\sc Directed Maximum Leaf}, as follows.

Let $a=0.526$. We can solve {\sc Directed Maximum Leaf} for a digraph $D$ on $n$ vertices using the following algorithm ADML:

\begin{description}
  \item[Stage 1.] Set $k:=\lceil an\rceil$. For each $x\in V(D)$ apply $\mathcal{B}(D,\{x\},\emptyset)$ to
  decide whether $D$ contains an out-branching with at least $k$ leaves. If $D$ contains such an out-branching,
  go to Stage 2. Otherwise, using binary search and $\mathcal{B}(D,\{x\},\emptyset)$,
  return the maximum integer $\ell$ for which $D$ contains an out-branching with $\ell$ leaves.
  \item[Stage 2.] Set $\ell:=\lceil an\rceil.$
  For $k=\ell+1,\ell+2,\ldots , n$, using Lemma \ref{OBexist}, decide whether $\hat{D}(\emptyset,S)$ has an out-branching for any vertex set $S$ of $D$
  of cardinality $k$ and if the answer is ``NO'', return $k-1$.
\end{description}

The correctness of ADML is obvious and we now evaluate its time complexity. Let $r=\lceil an\rceil$.
Since $3.72^a< 1.996$, Stage 1 takes time at most $3.72^rn^{O(1)}=O(1.996^n).$
Since $\frac{1}{a^a(1-a)^{1-a}}< 1.9973$, Stage 2 takes time at most $${n\choose r}\cdot n^{O(1)}=\left(\frac{1}{a^a(1-a)^{1-a}}\right)^nn^{O(1)}=O(1.9973^n).$$ Thus, we obtain the following:

\begin{theorem}
There is an algorithm to solve {\sc Directed Maximum Leaf} in time $O(1.9973^n).$
\end{theorem}

\section{Linear Kernel for {\sc Directed $k$-Leaf} restricted to Acyclic Digraphs}

Lemma \ref{OBexist} implies that an acyclic digraph $D$ has
an out-branching if and only if $D$ has a single vertex of in-degree zero. Since it is
easy to check that $D$ has a single vertex of in-degree zero, in what follows, we
assume that the acyclic digraph $D$ under consideration has a single
vertex $s$ of in-degree zero.

We start from the following simple lemma.

\begin{lemma}\label{OBDAG} In an acyclic digraph $H$ with a
single source $s$, every spanning subgraph of $H$, in which each
vertex apart from $s$ has in-degree 1, is an out-branching.\end{lemma}

Let $B$ be an undirected bipartite graph with vertex bipartition
$(V',V'')$. A subset $S$ of $V'$ is called a {\em bidomination set}
if for each $y\in V''$ there is an $x\in S$ such that $xy\in E(B).$
The so-called {\em greedy covering algorithm} \cite{asratian1998} proceeds as follows:
Start from the empty bidominating set $C$. While $V''\neq \emptyset$ do the following:
choose a vertex $v$ of $V'$ of maximum degree,
add $v$ to $C$, and delete $v$ from $V'$ and the neighbors of $v$ from $V''$.

The following lemma have been obtained independently by several
authors, see Proposition 10.1.1 in \cite{asratian1998}.

\begin{lemma}\label{bdgreedy}
If the minimum degree of a vertex in $V''$ is $d$, then
the greedy covering algorithm finds a bidominating set of size at most $1+\frac{|V_1|}{d}\left(1 + \ln \frac{d|V_2|}{|V_1|}\right).$
\end{lemma}

Let $D$ be an acyclic digraph with a single source. We use the
following reduction rules to get rid of some vertices of in-degree 1.

\begin{itemize}
\item[(A)] If $D$ has an arc $a=xy$ with $d^+(x)=d^-(y)=1$, then contract
$a$.
\item[(B)] If $D$ has an arc $a=xy$ with $d^+(x)\ge 2$, $d^-(y)=1$ and $x\neq s$,
then delete $x$ and add arc $uv$ for each $u\in N^-(x)$ and $v\in N^+(x)$.
\end{itemize}

The reduction rules are of interest due to the following:

\begin{lemma}\label{DD'-lem}
Let $D^*$ be the digraph obtained from an acyclic digraph $D$ with a
single source using Reduction Rules A and B as long as possible.
Then $D^*$ has a $k$-out-branching if and only if $D$
has one.
\end{lemma}
\begin{proof} Let $D$ have an arc $a=xy$ with $d^+(x)=d^-(y)=1$
and let $D'$ be the digraph obtained from $D$ by contracting $a$.
Let $T$ be a $k$-out-branching of $D$. Clearly, $T$ contains $a$ and let $T'$ be an out-branching
obtained from $T$ by contracting $a$.
Observe that $T'$ is also a $k$-out-branching whether $y$ is a leaf of $D$ or not.
Similarly, if $D'$ has a $k$-out-branching, then $D$ has one, too.

Let $D$ have an arc $a=xy$ with $d^+(x)\ge 2$, $d^-(y)=1$ and $x\neq s$
and let $D'$ be obtained from $D$ by applying Rule B. We will prove that $D'$ has a $k$-out-branching if and only if $D$ has one. Let $T$ be a $k$-out-branching in $D$. Clearly, $T$ contains arc $xy$ and $x$ is not a leaf of $T$. Let $U$ be the subset of $N^+(x)$ such that $xu\in A(T)$ for each $u\in U$ and let $v$ be the vertex such that $vx\in A(T)$. Then the out-branching $T'$ of $D'$ obtained from $T$ by deleting $x$ and adding arcs $vu$ for every $u\in U$ has at least $k$ leaves ($T'$ is an out-branching of $D'$ by Lemma \ref{OBDAG}). Similarly, if $D'$ has a $k$-out-branching, then $D$ has one, too.
\end{proof}

\vspace{2mm}

Now consider $D^*$. Let
$B$ be an undirected bipartite graph, with vertex bipartition
$(V',V'')$, where $V'$ is a copy of $V(D^*)$ and $V''$ is a copy of
$V(D^*)-\{s\}$. We have $E(B)=\{u'v'':\ u'\in V', v''\in V'', uv\in
A(D^*)\}.$

\begin{lemma}\label{D'B-lem}
Let $R$ be a bidominating set of $B$. Then $D^*$ has an
out-branching $T$ such that the copies of the leaves of $T$ in $V'$
form a superset of $V'-R$.
\end{lemma}
\begin{proof} Consider a subgraph $Q$ of $B$ obtained from $B$ by deleting all
edges apart from one edge between every vertex in $V''$ and its
neighbor in $R$. By Lemma \ref{OBDAG},
$Q$ corresponds to an out-branching $T$ of $D^*$ such that the
copies of the leaves of $T$ in $V'$ form a superset of $V'-R$.\end{proof}

\begin{theorem}
If $D^*$ has no $k$-out-branching, then the number $n^*$ of vertices in $D^*$ is less than $6.6(k+2).$
\end{theorem}
\begin{proof} Suppose that $n^*\ge 6.6(k+2)$; we will prove that $D^*$ has a $k$-out-branching.
Observe that by Rules A and B, all vertices of $D^*$ are of in-degree at least 2
apart from $s$ and some of its out-neighbors. Let $X$ denote the set of out-neighbors of $s$ of in-degree 1 and let $X''$ be the set of copies of $X$ in $V''$. Observe that the vertices of $V''-X''$ of $B-X''$ are all of degree at least 2. Thus, by Lemma \ref{bdgreedy},
$B-X''$ has a bidominating set $S$ of size at most $\frac{n^*}{2}(1+\ln 2)+1.$ Hence, $S\cup \{s\}$ is a bidominating set of $B$ and, by Lemma \ref{D'B-lem}, $D^*$ has a $b$-out-branching with $b\ge n^*-\frac{n^*}{2}(1+\ln 2)-2.$
It is not difficult to see that $b\ge \frac{n^*}{2}(1-\ln 2)-2\ge 0.153n^*-2\ge k.$\end{proof}

\section{Open Problems}

It would be interesting to see whether {\sc Directed $k$-Leaf} admits an
algorithm of significantly smaller running time, say $O(3^kn^{O(1)}).$
Another interesting and natural question is to check whether a linear-size
kernel exists for {\sc Rooted Directed $k$-Leaf} (for all digraphs).

\2

\noindent{\bf Acknowledgements}  Research of Gutin, Kim and Yeo
was supported in part by an EPSRC grant. Research of Daligault was supported in part by Alliance Project "Decomposition de graphes orient\'es" and ANR project GRAAL. We are thankful to Serge Gaspers for his ideas leading to Section \ref{admlsec}.


{\small
\begin{thebibliography}{10}

\bibitem{alonLNCS4596} N. Alon, F.V. Fomin, G. Gutin, M. Krivelevich, and S. Saurabh.
Parameterized Algorithms for Directed Maximum Leaf
Problems. {\em Proc. 34th ICALP},  Lect. Notes Comput. Sci. 4596 (2007), 352--362.

\bibitem{alonLNCS4855} N. Alon, F.V. Fomin, G. Gutin, M. Krivelevich, and S. Saurabh.
Better Algorithms and Bounds for Directed Maximum
Leaf Problems. {\em Proc. 27th Conf.
Foundations Software Technology and Theoretical Computer Science}, Lect. Notes Comput. Sci. 4855 (2007), 316--327.

\bibitem{alonSIAMJDM} N. Alon, F.V. Fomin, G. Gutin, M. Krivelevich and S. Saurabh,
Spanning directed trees with many leaves. SIAM J. Discrete Math. 23 (2009), 466--476.


\bibitem{asratian1998} A.S. Asratian, T.M.J. Denley, and R. H\"aggkvist,
{\em Bipartite Graphs and Their Applications}, Univ. Press, Cambridge,
1998.

\bibitem{bang2000} J. Bang-Jensen and G. Gutin, Digraphs: Theory,
Algorithms and Apllications, Springer-Verlag, London, 2000.

\bibitem{bodlaenderLNCS5125} H.L. Bodlaender, R.G. Downey, M.R. Fellows and D. Hermelin. On Problems without Polynomial Kernels. Lect. Notes
Comput. Sci. 5125 (2008), 563--574.

\bibitem{bonsmaLNCS2747} P.S. Bonsma, T. Brueggermann and G.J. Woeginger.
A faster FPT algorithm for finding spanning trees with many
leaves. Lect. Notes Comput. Sci. 2747 (2003), 259--268.

\bibitem{bonsma2007} P.S. Bonsma and F. Dorn. An FPT algorithm for directed spanning $k$-leaf.
Tech. Report (2007) \url{http://arxiv.org/abs/0711.4052}

\bibitem{bonsmaESA} P.S. Bonsma and F. Dorn. Tight bounds and
faster algorithms for Directed Max-Leaf. {\em Proc. 16th ESA}, Lect. Notes Comput. Sci.
5193 (2008), 222--233.

\bibitem{dingJGT3} G. Ding, Th. Johnson, and P. Seymour. Spanning trees with many leaves.
J. Graph Theory 37 (2001), 189--197.

\bibitem{downey1999} R.G. Downey and M.R.~Fellows,
{\em Parameterized Complexity\/}, Springer, 1999.

\bibitem{drescherACMTA} M. Drescher and A. Vetta. An approximation algorithm for the maximum leaf spanning arborescence problem. To appear in ACM Transactions on Algorithms.

\bibitem{estivillACID05} V. Estivill-Castro, M.R. Fellows, M.A. Langston, and F.A. Rosamond,
FPT is P-Time Extremal Structure I. Proc. ACiD'05, College Publications, London (2005), 1--41.

\bibitem{fellowsLNCS1974} M.R. Fellows, C. McCartin, F.A. Rosamond, and U. Stege. Coordinated kernels and catalytic reductions: An improved FPT algorithm for max leaf spanning tree and other problems. Lect. Notes Comput. Sci. 1974 (2000), 240--251.

\bibitem{fernau} H. Fernau, F.V. Fomin, D. Lokshtanov, D. Raible, S. Saurabh, and Y. Villanger,
Kernel(s) for problems with no kernel: on out-trees with many leaves. Tech. Report (2008) \url{http://arxiv.org/abs/0810.4796v2}

\bibitem{flum2006} J. Flum and M. Grohe, {\em  Parameterized Complexity
Theory}, Springer, 2006.

\bibitem{fominA52} F.V. Fomin, F. Grandoni and D. Kratsch. Solving Connected Dominating Set Faster Than $2^n$. Algorithmica 52 (2008), 153--166.

\bibitem{galbiatiTCS181} G. Galbiati, A. Morzenti, and F. Maffioli. On the approximability of some maximum spanning tree problems. Theor. Computer Sci. 181 (1997), 107--118.

\bibitem{griggsDM104} J. R. Griggs and M. Wu. Spanning trees in graphs of minimum degree four or five.
Discrete Math. 104 (1992), 167--183.

\bibitem{kleitmanSIAMJDM4} D.J. Kleitman and D.B. West. Spanning trees with many leaves. SIAM J. Discrete Math. 4 (1991), 99--106.

\bibitem{kneisISAAC2008} J. Kneis, A. Langer and P. Rossmanith. A new algorithm for finding trees with many leaves. {\em Proc. ISAAC 2008}, Lect. Notes Comput. Sci. 5369 (2008), 270--281.

\bibitem{linial1987} N. Linial and D. Sturtevant. Unpublished result (1987).

\bibitem{luJA29} H.I. Lu and R. Ravi. Approximating maximum leaf spanning trees in almost linear time. J. Algorithms 29 (1998), 132--141.

\bibitem{niedermeier2006}
R.~Niedermeier. {\em Invitation to Fixed-Parameter Algorithms},
Oxford University Press, 2006.

\bibitem{solisLNCS1461} R. Solis-Oba. 2-approximation algorithm for finding a spanning tree with the maximum number of leaves. Lect. Notes Comput. Sci. 1461 (1998), 441--452.


\end{thebibliography}
}
\end{document}